\documentclass[superscriptaddress,preprint]{revtex4}
\usepackage{times}
\usepackage{amsmath,amssymb}
\usepackage{graphicx,subfigure}
\usepackage{bm}
\usepackage{psfrag}
\usepackage{slashbox}
\usepackage{relsize}

\newcommand\Cpp{C\nolinebreak[4]\hspace{-.05em}\raisebox{.4ex}{\relsize{-3}{\bf ++}}}

\begin{document}

\title{DynamO: A free ${\mathcal O}(N)$ general event-driven
  molecular-dynamics simulator}

\author{M.~N.~Bannerman}
\affiliation{Institute for Multiscale Simulation, Universit\"at
  Erlangen-N\"urnberg, Erlangen, Germany}
\affiliation{School of Chemical Engineering and Analytical Science,
  The University of Manchester, Oxford Road, 
  Manchester M13 9PL, United Kingdom }

\author{R.~Sargant} 
\affiliation{School of Chemical Engineering and Analytical Science,
  The University of Manchester, Oxford Road, 
  Manchester M13 9PL, United Kingdom }

\author{L.~Lue}
\affiliation{School of Chemical Engineering and Analytical Science,
  The University of Manchester, Manchester, United Kingdom }
\affiliation{Department of Chemical and Process Engineering,
  University of Strathclyde, James Weir Building, 75 Montrose Street,
  Glasgow G1 1XJ, United Kingdom }


\date{Received: date / Accepted: date}

\begin{abstract}
  Molecular-dynamics algorithms for systems of particles interacting
  through discrete or ``hard'' potentials are fundamentally different
  to the methods for continuous or ``soft'' potential systems.
  Although many software packages have been developed for continuous
  potential systems, software for discrete potential systems based on
  event-driven algorithms are relatively scarce and specialized.
  We present DynamO, a general event-driven simulation package which
  displays the optimal ${\mathcal O}(N)$ asymptotic scaling of the
  computational cost with the number of particles $N$, rather than the
  ${\mathcal O}(N\log N)$ scaling found in most standard algorithms.
  DynamO provides reference implementations of the best available
  event-driven algorithms.  These techniques allow the rapid
  simulation of both complex and large ($>10^6$ particles) systems for
  long times.
  The performance of the program is benchmarked for elastic hard
  sphere systems, homogeneous cooling and sheared inelastic hard
  spheres, and equilibrium Lennard-Jones fluids.
  This software and its documentation are distributed under the GNU
  General Public license and can be freely downloaded from \newline
  {\tt http://marcusbannerman.co.uk/dynamo}.

  \keywords{molecular dynamics \and event-driven simulation \and discontinuous potentials \and hard spheres \and square-well potential}
\end{abstract}
\maketitle

\section{Introduction}

Molecular-dynamics simulations have become an indispensable tool in
the development of novel nanomaterials~\cite{Fish_2006}, drug
discovery~\cite{Hansson_etal_2002,Wong_McCammon_2003,Jorgensen_2004},
and materials engineering in the estimation of thermophysical
properties and phase behavior of complex solutions.  Molecular
dynamics (MD) not only allows the exploration of the link between
inter-particle interactions and macroscopic structure and dynamics,
but it is also capable of providing quantitative predictions for real
materials.
Molecular-dynamics simulations have been dominated by time stepping
methods for systems that interact with continuous potentials.  This
method was first used by Rahman~\cite{Rahman_1964} in 1964, and later
popularized by Verlet~\cite{Verlet_1967}.
Since then, many sophisticated software packages have been developed,
such as LAMMPS (Large-scale Atomic/Molecular Massively Parallel
Simulator)~\cite{Plimpton_1995}, GROMACS (Groningen Machine for
Chemical Simulation)~\cite{gromacs_2005,gromacs_4.0},
NAMD~\cite{Phillips_etal_2005}, Desmond~\cite{Bowers_etal_2006} and
ESPResSo (Extensible Simulation Package for Research on Soft
Matter)~\cite{Limbach_etal_2006}, which are freely available and allow
the simulation of complex systems.
In addition, a wealth of force fields based on continuous potentials
have been developed to describe real materials, such as small organic
and inorganic molecules, polypeptides, proteins, and DNA (e.g., see
AMBER~\cite{Cheatham_Young_2001,Ponder_Case_2003} or
CHARMM~\cite{CHARMMFF}).  MD has also been applied to granular
systems, which was pioneered by Cundall and
Strack~\cite{Cundall_Strack_1979}.  Since then, Hertz's law for
elastic particles has been generalized for viscoelastic
spheres~\cite{Brillantov_etal_1996}, and a range of approximations for
the tangential forces are now
available~\cite{Deltour_Barrat_1997,Haff_Werner_1986,Walton_Braun_1986a}.


An alternative approach to modeling many-body systems is through the
use of discrete interaction potentials, such as the hard-sphere or
square-well potentials.  These potentials contain only distinct energy
level changes; however, they can be stepped to either approximate soft
potentials, such as the Lennard-Jones
potential~\cite{Chapela_etal_1989}, or directly reproduce
thermodynamic data~\cite{Elliot_2002}.

The ``true'' interactions between real molecules or atoms (or larger
scale particles) are expected to be smooth and continuous, and so one
may question the relevance of discrete potentials; they are an extreme
approximation to the ``true'' interactions.  However, in reality, all
interaction potentials that are used in computer simulations are
necessarily approximate, due to practical limitations in computing
resources.  In the case of continuous potentials, the main
approximation is typically the assumption of pair-wise additivity,
neglecting many-body interactions which are present in all ``real''
systems, or the use of restricted functional forms for the interaction
potential (e.g., Lennard-Jones, Stockmayer, etc.).  So most commonly
used continuous potentials also only approximate the ``true''
interactions.

The main issue is, however, not whether a potential exactly reproduces
the ``true'' interactions between real particles, but whether it
captures the essential features of the interaction to be able to
reproduce the physics/chemistry which is of interest in a particular
study.  This is the primary motivation of coarse grained simulations,
and many coarse grained potentials have also been developed to
describe systems on larger scales, such as the M3B model for
carbohydrates \cite{Molinero_Goddard_2004} and the MARTINI force
field, which was originally developed for
lipids~\cite{Marrink_etal_2007} and later extended to
proteins~\cite{Monticelli_etal_2008}.
It is within this context that discrete potentials are extremely
useful.

Although simulations for discrete potential systems are not as
prevalent as for continuous potential systems, the literature for
classical systems of particles with discontinuous potentials is quite
extensive.  Indeed, Alder et al.~\cite{ALDER_WAINWRIGHT_1957} reported
molecular-dynamics simulations for systems of hard spheres using an
event-driven algorithm in 1957, seven years before Rahman's soft
potential simulations.  Since then, many discrete potential force
fields have been developed for a wide range of systems, including
granular materials~\cite{Goldhirsch_Zanetti_1993}, simple molecular
systems (e.g., SPEADMD~\cite{Unlu_etal_2004}) such as mixtures of
hydrocarbons~\cite{Unlu_etal_2004}, ethers~\cite{Unlu_etal_2004},
alcohols~\cite{Elliott_etal_2007}, amines~\cite{Elliott_etal_2007},
and carboxylic acids~\cite{Vahid_Elliott_2010}, block copolymer
micelles~\cite{Woodhead_Hall_2010} and organized
mesophases~\cite{Schultz_etal_2002}, and detailed models (e.g.,
PRIME~\cite{Nguyen_Hall_2004}) for
polypeptide~\cite{Hall_Wagoner_2006,Marchut_Hall_2007} and
protein~\cite{Cheon_etal_2010} solutions.
%
%
These models do not just offer qualitative insight to these systems,
but they also provide quantitative predictions for such properties as
vapor-liquid phase equilibrium~\cite{Unlu_etal_2004}.  In addition,
the models are detailed enough to realistically capture protein
structure, but sufficiently efficient to examine folding and
aggregation~\cite{Nguyen_Hall_2004,Cheon_etal_2010}.

In coarse grained simulations, discrete potentials possess an
advantage over continuous potentials in their greater simplicity.  The
methods used to simulate discrete potential systems offer great
computational advantages at low to moderate densities and allow
computational resources to be focused on the regions in space and time
where relevant dynamics occurs.  Consequently, much larger length and
time scales can be explored for discontinuous coarse-grained
potentials than equivalent coarse-grained models based on continuous
potentials.  A comparison of relative advantages of discrete and
continuous potential models, within the context of fibril formation in
polypeptides, is provided in Ref.~\cite{Hall_Wagoner_2006}.

Standard numerical methods developed for integrating systems
interacting through soft potentials are inefficient for systems with
discrete potentials, due to discontinuities in the potential.  A
time-stepping algorithm, where changes in the interaction energies are
detected after the time step is taken, is still
feasible~\cite{Allen_etal_1989}; however, this method is necessarily
approximate, and high accuracy requires a small,
computationally-expensive time-step.
Event-driven algorithms avoid these difficulties by detecting the time
of the next interaction {\em a priori}. The system is then
analytically integrated to the time of the next interaction (event) in
a single step.  In principle, event-driven algorithms provide an exact
method for performing molecular-dynamics simulations for discrete
potential systems, where the dynamics can be decomposed into a
sequence of events.  Event-driven MD is extremely efficient for
simulating systems of particles interacting through steep potentials
with a relatively small number of steps.

There are numerous reviews on algorithms for event-driven molecular
dynamics~\cite{Allen_etal_1989,ALDER_WAINWRIGHT_1959,Rapaport_1980,Lubachevsky_1991,Marin_etal_1993,Sigurgeirsson_2001,Donev_2009};
however, it is difficult to find documented implementations of 
algorithms which include the source code.  Modern algorithms are
complex and contain many subtle difficulties which, if poorly
implemented, can severely restrict the generality and speed of the
code.
In this article, we present DynamO (DYNAMics of discrete Objects), a
free source molecular-dynamics package that is optimized for
event-driven dynamics.  DynamO is capable of simulating large
($\gtrsim10^6$ particles) and complex systems for extremely long
simulation times~\cite{Bannerman_etal_2010}.  This package has already
been used to study sheared/damping granular
materials~\cite{Bannerman_etal_2009,Bannerman_etal_2011}, square well
molecules~\cite{Bannerman_Lue_2010}, binary
mixtures~\cite{Bannerman_Lue_2009}, parallel
cubes~\cite{Hoover_etal_2009}, and helix forming
polymers~\cite{Bannerman_etal_2009_2}.

The remainder of this article is organized as follows.
In Section~\ref{sec:algorithm}, we begin with an overview of the basic
elements of event-driven molecular-dynamics simulations.  This section
briefly reviews some of the recent algorithmic advances.
Some improvements that we have developed to the simulation algorithm
are presented in Section~\ref{sec:innovations}.  These include novel
approaches to access particle information, optimize neighbor lists,
and minimize numerical inaccuracies.
%
%
Section~\ref{sec:timing} provides benchmark simulations for DynamO on
systems of single component hard spheres and stepped Lennard-Jones
molecules.  These simulations provide timing results and information
on the scaling of the calculation times with system size.
Finally, the conclusions of the paper are presented in
Section~\ref{sec:conclusions}, along with a discussion of possible
directions for future extensions to DynamO.  
An outline of the general structure of DynamO and details of various
aspects of its implementation, as well as a listing of its currently
implemented features, are provided in Appendix~\ref{sec:programdesign}.

\section{Algorithm details\label{sec:algorithm}}

A molecular-dynamics simulation calculates the trajectory of a
collection of a large number of interacting particles.  When the
particles interact with each other through a discrete potential, the
dynamics of the system are governed by a series of distinct events
(e.g., collisions between particles).  These events may alter the
properties of the particles, such as their velocities.  Between these
events, the particles move on a ballistic trajectory, and the dynamics
of the system is known analytically. This is a significant advantage
of event-driven algorithms, as the ballistic motion requires no
numerical integration and does not suffer from truncation error.
%

In an event-driven simulation, there are three major tasks which
occupy most of the computational time: (i) searching for events (event
detection), (ii) maintenance of the event list, and (iii) execution of
events.  An outline of a basic event-driven simulation algorithm is
given below, including the scaling of the computational cost of each step
with the number of particles $N$ in the system:

\begin{enumerate}
\item \label{en:eventtest} Event testing ${\mathcal O}(N^2)$: All particles and
  pairs of particles are tested to determine if/when the next
  interaction occurs.  The times of these events are inserted into the
  Future Event List (FEL).

\item \label{en:eventsort} Event sorting ${\mathcal O}(N)$: The events
  in the FEL are sorted to determine the next event to occur.

\item \label{en:freestream} Motion of the system ${\mathcal O}(N)$:
  The system is evolved, or free streamed, to the time of the next
  event.

\item \label{en:eventexecution} Execution of the event ${\mathcal
    O}(1)$: Particles involved in the event are updated with new
  velocities.

\item \label{en:updateFEL} Update events in FEL ${\mathcal O}(N)$:
  Events in the FEL that involve particles that have just undergone
  the executed event are now invalid.  New events for these particles
  are possible and must be tested for.  The future event list may be
  cleared and rebuilt (${\mathcal O}(N^2)$), or only the affected
  events can be updated (${\mathcal O}(N)$).

\item \label{en:endcondition} End condition: Continue to
  step~\ref{en:eventsort} unless sufficient collisions have been
  executed or the maximum simulation time has elapsed.
\end{enumerate}

Alder and coworkers~\cite{ALDER_WAINWRIGHT_1957} proposed the first
algorithm for event-driven molecular-dynamics simulations.
Since their pioneering work, there have been significant advances in
the development of this initial algorithm.
We will briefly cover the advances in event detection/execution before
detailing improvements in the maintenance of the list of all possible
future events.

Alder and Wainwright~\cite{ALDER_WAINWRIGHT_1959} were the first to
suggest the use of a ``neighbor list'' technique to improve the
efficiency of the search for future events.  In this method, the
simulation box is divided into small cells. An additional event type
is also introduced to track the motion of the particles between these
cells. These cells are used to determine which particles are close to
a particle undergoing an event. This significantly reduces the number
of particles that need to be tested for new interactions in
step~\ref{en:eventtest} to ${\mathcal O}(N)$ and in
step~\ref{en:updateFEL} to ${\mathcal O}(1)$.
%
This ``neighbor list'' technique has since been extended to infinite
systems~\cite{Marin_Cordero_1996} using a hashing technique,
overlapping cells~\cite{Krantz_1996} to reduce the frequency of cell
transitions, and hybrid methods~\cite{Vrabecz_Toacuteth_2006}
combining multiple neighbor lists. All of these methods are available
in DynamO.

One of the most significant advances in event-driven simulation came
with the development of
asynchronous~\cite{Jefferson_1985,Lubachevsky_1991} or ``delayed
states''~\cite{Marin_etal_1993} algorithms. In these algorithms, only
particles involved in an event or within the neighborhood of an event
are updated when an event occurs (step~\ref{en:freestream}).  Each
particle stores the time at which it was last updated and is only
evolved to the current time when it is tested for, or undergoes an
event.  As particles which are not involved in an event incur no
overhead, step~\ref{en:freestream} is reduced to ${\mathcal O}(1)$.

In most modern implementations of event-driven algorithms, the system
size scaling of the event computational cost arises from the
maintenance~\cite{Rapaport_1980} of a list of all possible future
events (step~\ref{en:eventsort}). This list needs to be sorted to
determine the next event to occur, and after an event is executed the
list must be updated.  Improvements in sorting began with the use of
many variants of binary trees~\cite{Rapaport_1980} before complete
binary trees were found to be optimal~\cite{Marin_Cordero_1995}.
Complete binary trees exhibit an ${\mathcal O}(\log_2N_{\rm event})$
scaling in the number of events $N_{\rm{}event}$ contained in the
tree.  It has been previously
suggested~\cite{Rapaport_1980,Lubachevsky_1991} that an ${\mathcal
  O}(\log_2N_{\rm{}event})$ scaling of sorting the event list is the
asymptotic minimum; however, a recent advance in event sorting using
calendar event queues~\cite{Paul_2007} now enables ${\mathcal O}(1)$
scaling.  This is achieved by first presorting events into fixed
intervals of time or ``dates'' within a ``calender''. The date to
which an event corresponds can be determined by simply dividing the
event time by the duration of a date and truncating to an integer
(${\mathcal O}(1)$).  As the simulation progresses to a new date in
the calendar, the events within a date are sorted and processed using
a complete binary tree.  The length and duration of the calendar is
scaled with the system size, resulting in a fixed average size of this
complete binary tree.  This ensures that the deletion, insertion, and
updating operations on the FEL all remain of order ${\mathcal O}(1)$.

With these methods, the overall computational cost of executing a
single event is now independent of the system size, which is the
theoretical optimum.  The number of events per particle is typically
proportional to the total time simulated, thus the computational cost
of simulating a unit of time scales as ${\mathcal O}(N)$ with the
system size.

Other algorithmic improvements, which do not affect the system size
scaling of the computational cost, have been made to the deletion of
events from the calendar.  The local minima
algorithm~\cite{Marin_etal_1993} relies on each event being associated
with at least one particle.  A priority queue called a particle event
list (PEL) is used to sort the events associated with a single
particle.  The PEL is then inserted into the global event list and
sorted according to its earliest event.  When a particle undergoes an
event, at least half of the invalidated events associated with it can
be deleted by simply erasing the corresponding PEL.  The remainder of
the invalidated events are left in the FEL and are deleted if they
reach the top of the FEL.  This is achieved by tracking the number of
events each particle has undergone in total and the value of this when
the event was tested~\cite{Marin_etal_1993}.

\section{Improvements to the simulation method
\label{sec:innovations}}

In this section, we present some additional improvements to the
event-driven simulation algorithm that we have developed and
implemented within DynamO.  These methods are primarily concerned with
the storage of data and increasing simulation accuracy.

\subsection{Particle data}

The main bottleneck in most scientific programs is the speed of memory
accesses, and event-driven simulation is no exception. It is typically
cheaper to calculate values than to store them in memory. In DynamO,
this approach is applied to all static values or ``properties''.
Associated to each particle is a class which contains only a
particle's position, velocity, ID number, and the time the particle
was last updated.  All other properties, dependent on the system
studied (e.g., mass/inertia, orientation, species), are accessed using
the particle's ID number when required.

A design feature of DynamO is the use of functions, as opposed to
look-up tables, to perform this look-up both when mapping a property
(such as mass) to a range of particles and when determining the values
of a property. For example, in a single component system all particles
have identical masses. To avoid storing redundant information, a
distribution representing a single value of the mass is stored behind
a function that maps it to all particles. These functions behave like
a standard STL-container and are used to define molecular topology,
species, mass, and how particles interact (e.g., to create mixtures of
particles).  This approach conserves memory, as an absolute minimal
number of entries are required, and is faster than a look-up table due
to the reduced number of memory accesses.

For dynamical properties of particles where this approach is
impossible (e.g., if two particles have captured each other in an
attractive well), DynamO makes use of unordered sets and maps. These
hashed containers still provide ${\mathcal O}(1)$ operations when
using a suitable hash function but conserve memory when compared to
using arrays.

\subsection{Morton ordered neighbor lists
\label{sec:MortonOrdering}}

The algorithms discussed in Sec.~\ref{sec:algorithm} result in a
theoretical system size scaling of ${\mathcal O}(1)$ in the cost of
processing a single event.  In practice, the computational cost is
affected by the memory architecture of the computer that runs the
simulation.  If the events are relatively inexpensive to test for, the
bulk of the simulation time will be spent on memory accesses to
retrieve particle, event, and neighbor list data.

A fundamental aspect of modern processors is the use of a CPU cache,
into which data are ``fetched'' before becoming available to the
running process.  A cache ``miss'' occurs when data to be accessed are
not already available in the cache.  The cost of a cache miss is
typically quite severe, requiring several computational cycles to
fetch the data from main memory and load it into the CPU cache.  Data
are typically fetched in blocks of $64$ bytes; therefore, data
localized in memory are typically fetched at the same time.  Thus, if
the location of data in memory is strongly correlated to the data
access pattern, then the number of cache misses can be reduced.

The particle and event data accesses are effectively random, which
renders any attempt to optimize the access patterns futile; however,
accesses to the neighbor list data are strongly correlated.  Whenever
a particle's local space is to be inspected for possible events, the
cells of the neighbor list which surround the particle are checked for
event partners.  In the standard implementation of a neighbor
list~\cite{ALLEN_TILDESLEY_1987}, only the first element of each cell
is correlated this way as a singly linked list is used look-up all
other contained particles. Nevertheless, accesses to the first
particle in the neighbor list cell are strongly correlated in the
spatial coordinates of the cell. Thus, if the cell's first particle
data are arranged such that spatially localized data are also
localized in memory, the number of cache misses will be reduced.

\begin{figure}
\centering
\includegraphics[clip,width=0.8\columnwidth]{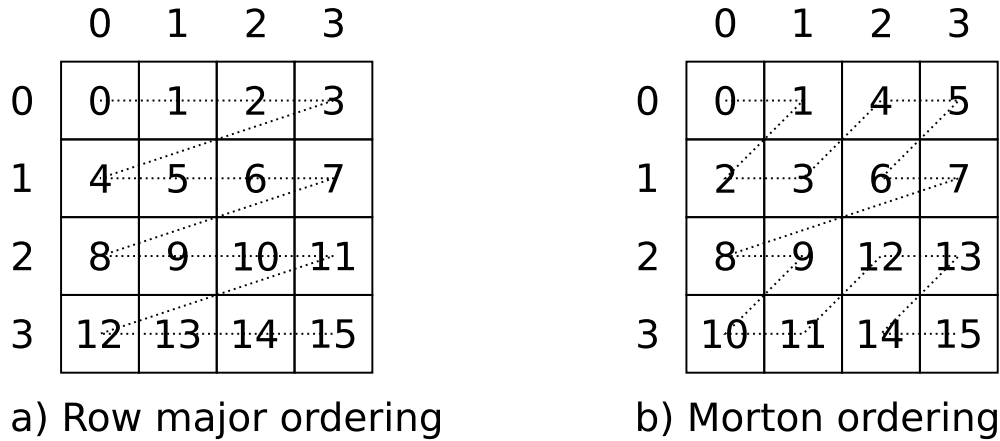}
\caption{\label{fig:morton} A two dimensional $4\times4$ array stored
  in row major order and Morton order in memory.}
\end{figure}
Arrays are typically stored linearly (row major in \Cpp) in computer
memory, where each successive row of data in the lowest spatial
dimension is appended to the previous row (see
Fig.~\ref{fig:morton}a).  An alternative space filling curve which
retains a high level of spatial locality is the Morton-order or
``Z-order'' curve (see Fig.~\ref{fig:morton}b).  Recently, fast
methods for dilating integers used in Morton
ordering~\cite{Raman_Wise_2008} and methods for directly carrying out
mathematical operations on the dilated integers~\cite{Adams_Wise_2006}
have become available.  The overhead of calculating a three
dimensional Morton number is now less than the cost of a cache miss in
many applications.  In DynamO, we have implemented Morton ordering in
the neighbor list, and a comparison between linear and Morton order is
presented in the timing results section (see
Section~\ref{sec:timing}).

The algorithms described thus far are primarily concerned with
increasing the speed of the calculations; however, the numerical
accuracy of the simulation is crucial. The methods employed to ensure
precision is maintained are detailed in the following section.

\subsection{Min-Max Particle Event Lists\label{sec:MinMax}}
Priority queues are often used for the PEL (Particle Event List) as
they allow insertion, access to the shortest time event, and clearing;
and are optimal for the access patterns of the
PEL~\cite{Poschel_Schwager_2005}. However, there are several drawbacks
to using these containers which result from their ability to have a
dynamic size: Often only the first few events in the PEL are relevant
to the dynamics and yet all tested events are stored in the priority
queue. Additionally, the methods for manipulating dynamic memory ({\tt
  new} and {\tt delete} in \Cpp) are slow and result in an extra layer
of indirection.  One method proposed to solve these drawbacks is to
only store the earliest event in a particle's event list. If this
single stored event is invalidated, all possible events of the
particle must be recalculated and the new minimum stored
~\cite{Lubachevsky_1991}. Unfortunately, general event-driven
simulations utilize many types of virtual events (e.g., neighbor list
boundary crossings), resulting in many recalculations. Modern event
driven potentials, such as those for asymmetric particles, can also be
expensive to recompute.

It would be preferable to allow the PEL to only store some small, but
greater than one, number of events.  This would limit the number of
events in the PEL (to save memory) and yet still store a sufficient
number to minimize recalculations which occur when the PEL is
empty. This fixed-size PEL must provide constant-time access to the
earliest event (for the dynamics), and also provide constant-time
access to the last event in the list to mark it as a
full-recalculation event if an event is added to a filled PEL. This
last element may also be used as a early test mechanism to determine
if an event needs to be inserted into a filled PEL.
\begin{figure*}[tb]
\begin{center}
  \includegraphics[width=\columnwidth,clip]{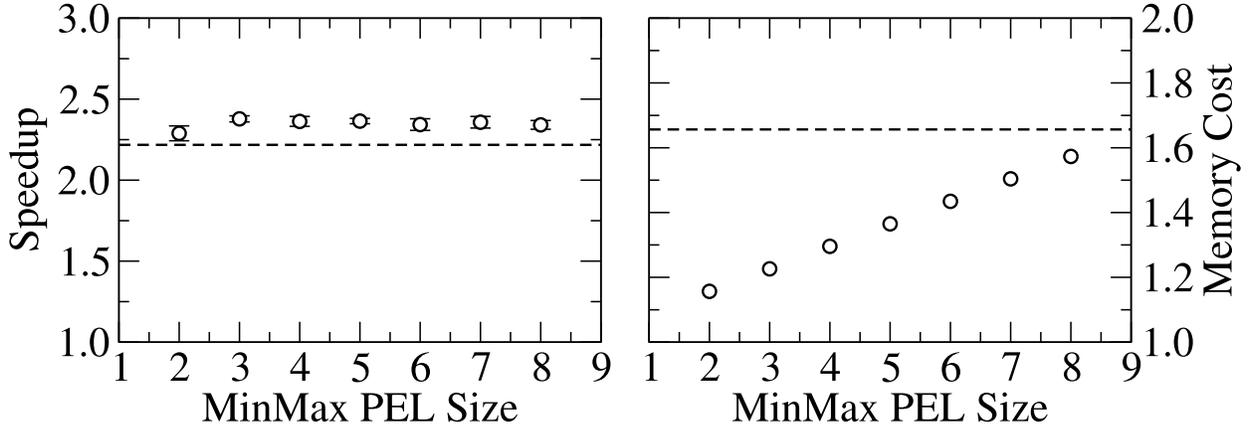}%
\caption{\label{fig:PELcomparison}
  The speedup and memory cost of using a MinMax PEL of various sizes
  relative to storing only a single event per particle. The dashed
  lines are the values for using a STL priority\_queue. The results
  are for a monodisperse system of $N=10^5$ elastic hard spheres
  simulated on an Intel\textregistered\ Core\texttrademark i5 750 with
  8~MB L3 cache and 8~GB of RAM.}
\end{center}
\end{figure*}

MinMax heaps~\cite{Atkinson_etal_1986} are a data structure which
satisfies all of these requirements.  A comparison of MinMax heaps,
single event storage, and a standard template library (STL) {\tt
  priority\_queue} is presented in Fig.~\ref{fig:PELcomparison}.
Even for hard sphere systems, the MinMax PEL offers the speed
advantage of a {\tt priority\_queue} with a great saving in memory
cost. A slight speed improvement from the {\tt priority\_queue} is
expected, as the MinMax PEL does not require dynamic memory. The
optimal size of a MinMax queue appears to be 3 stored events. This
optimum is dependent on how many invalid or virtual events are
expected to appear at the top of the PEL; however, a more conservative
size of 4 or 5 may be selected without too great an increase in the
memory cost. All further simulation results presented in this paper
are performed with a MinMax PEL size of 4.

\subsection{Accuracy and Time Invariance
\label{sec:accuracy}}

Molecular-dynamics simulations require a high level of accuracy.  This
is especially true for ``hard core systems'', where configurations
with overlapping hard cores resulting from numerical inaccuracies are
unphysical and impossible to resolve.  Accuracy is particularly
important in inelastic, granular systems due to clustering of the
particles; small errors in the movement of particles, resulting in
overlaps, are increasingly probable.

Several improvements have been made to the algorithm used in DynamO to
maintain the numerical precision of the simulation.  The use of the
``delayed states''~\cite{Marin_etal_1993} algorithm already reduces
the number of times a particle is free-streamed between events.  Also,
when an event is scheduled to occur, events are retested to determine
the exact time at which it occurs~\cite{Poschel_Schwager_2005}.  This
reduces the likelihood of overlaps occurring due to inaccuracies in
the free streaming.  Furthermore, the dynamics of the system is
tracked to ensure invalid events cannot occur (e.g., particles must be
approaching to be tested for a collision, square-well molecules must
have been captured to test for a release event).  This implies that
the dynamics of the system must always be deterministic; however,
random events, such as those that occur in the Andersen
thermostat~\cite{Andersen_1980}, are possible by randomly assigning a
time after each random event and scheduling this fixed time in the
event list.

Another inaccuracy arises from the storage of absolute times in the
simulation.  Events occur at an absolute time $t_{\rm e}$ and, within
the asynchronous algorithms, the particle data are stored at a certain
absolute simulation time $t_{\rm p}$.  As the absolute simulation time
$t_{\rm sim}$ increases in magnitude, round-off error will accumulate
in these stored absolute times.  To avoid this, only time differences
$\Delta{}t_{\rm e/p}$, recording the time relative to the current
absolute simulation time
$t_{\rm{}e/p}=t_{\rm{}sim}+\Delta{}t_{\rm{}e/p}$, must be stored.
This does, however, introduce an ${\mathcal O}(N)$ computational cost
per event in maintaining these time differences.  This is alleviated
by storing time differences relative to a reference time difference
$\Delta{}t_{\rm{}e/p}=\Delta{}t_{\rm{}ref}+\Delta{}t_{\rm{}e/p}^{\rm
  (stored)}$.  This reference time difference $\Delta t_{\rm ref}$ is
updated at every event without incurring significant cost.
Periodically all time differences and reference time differences must
be synchronized to prevent round-off error, and the interval at which
this synchronization occurs is proportional to the system size.  This
leads to a time invariant simulation algorithm with a fixed upper
bound on round-off error {\em without} affecting the scaling of the
computational cost.

\subsection{Visualization}

\begin{figure*}[tb]
\begin{center}
  \includegraphics[width=0.5\columnwidth,clip]{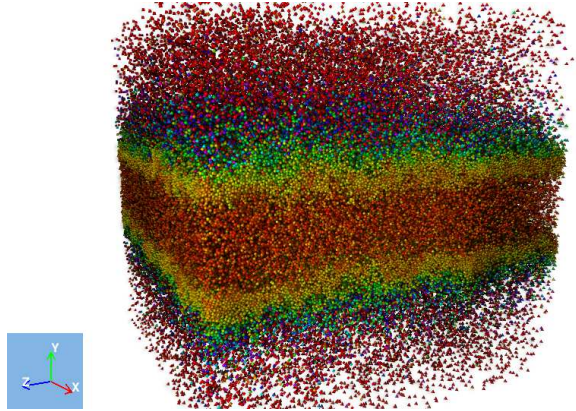}%
\caption{\label{fig:CoilSnapshot}
  A snapshot of $5\times10^5$ sheared inelastic hard spheres,
  generated live from a simulation, displaying a characteristic
  clustering behavior of granular systems. On a NVIDIA\textregistered\ 
  Geforce GTX 260\texttrademark, the image is updated at 15 frames per
  second. }
\end{center}
\end{figure*}

DynamO is capable of simulating millions of particles at close to real
time. With macroscopic simulations of granular systems, the simulation
speed approaches the timescale of the interesting dynamics. To enable
live visualization of these massive systems and to allow interactive
or ``steered'' simulations, a new visualizer was written in
OpenCL/GL. This library, known as Coil, is capable of rendering up to
a million spheres in real time with full diffusive, specular and
shadow lighting calculations and HDR effects. The result is
publication-quality images (see Fig.~\ref{fig:CoilSnapshot}) at
interactive frame-rates while only using a single core of the host
CPU. This visualization library is already finding application in a
wide range of simulators outside of DynamO.

\subsection{Summary}

In this section, we have discussed some of the new methods we have
developed to improve the computational efficiency of DynamO.
The next section provides some benchmarking simulations for DynamO and
tests the system size scaling of the simulation code.

\section{Benchmarking
\label{sec:timing}}

In order to benchmark the speed of DynamO and test that the optimal
${\mathcal O}(1)$ scaling is achieved, we perform molecular-dynamics
simulations on systems composed of hard spheres.  This interaction is
relatively inexpensive and as such is useful in testing the
performance of the simulation framework.  Each sphere has a diameter
$\sigma$ and is run over a range of reduced number densities
$\rho\sigma^3$. The simulations were performed on a desktop computer
with an Intel\textregistered Core\texttrademark i5 750 processor with
8~GB of RAM.  The simulations utilized only a single core of the
processor and are averaged over $4$ runs of $5\times10^6$
collisions. The optimal parameters for the bounded priority queue are
determined at the start of the simulation by instrumenting the initial
event distribution. $N$ calendar dates are used with a width equal to
the mean time between events.

\begin{figure*}[tb]
\begin{center}
  \includegraphics[width=\columnwidth,clip]{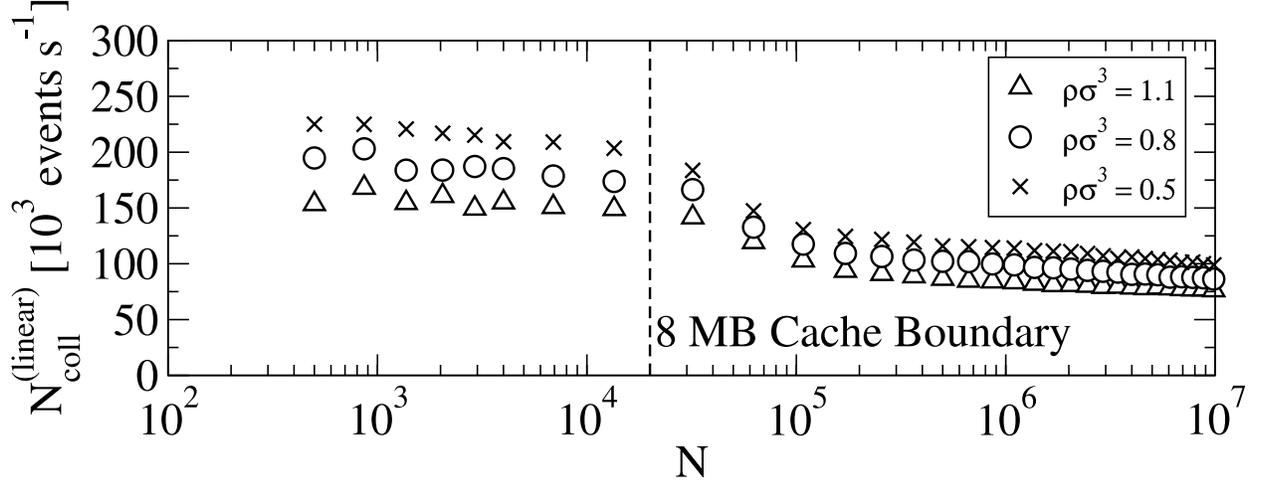}%
\caption{\label{fig:scaling}
  The number of collision events processed per second using linear
  neighbor lists $N_{\rm coll}^{(\rm linear)}$ as a function of (a)
  the number of particles $N$ and (b) the number density $\rho$.  The
  lines indicating the cache memory boundary is approximate as
  different densities incur slightly different memory requirements.
  The results are for a monodisperse system of elastic hard spheres
  simulated on an Intel\textregistered Core\texttrademark i5 750 with
  8~MB L3 cache and 8~GB of RAM.}
\end{center}
\end{figure*}

The average number of collisions per second is plotted in
Fig.~\ref{fig:scaling}a as a function of system size and density.  It
is apparent that the memory architecture plays a large role in the
speed of the simulation~\cite{Paul_2007}.  The rate of collision
maintains a relatively constant value when the program fits inside the
CPU cache ($\le 8$~MB boundary, $N\lesssim1.6\times10^4$), and for
very large systems ($N\gtrsim 10^5$) where the cache effects are
proportionally small. Accounting for these memory size effects, the
algorithm appears to exhibit ${\mathcal O}(1)$ scaling of the
collision cost. Inside the cache, the simulation reaches a maximum of
roughly $2.3\times 10^5$ events per second, compared to a minimum of
approximately $7\times10^4$ events per second outside of the cache.
Beyond the $8$~GB memory limit ($N\approx1.6\times10^7$) disk swapping
begins to occur and the performance is substantially degraded. The
event processing rate is relatively insensitive to density.  A slight
decrease in the event processing rate is expected at higher densities
as local neighbor lists contain more entries (neighbors).

\begin{figure}[tb]
\begin{center}
\includegraphics[width=\columnwidth,clip]{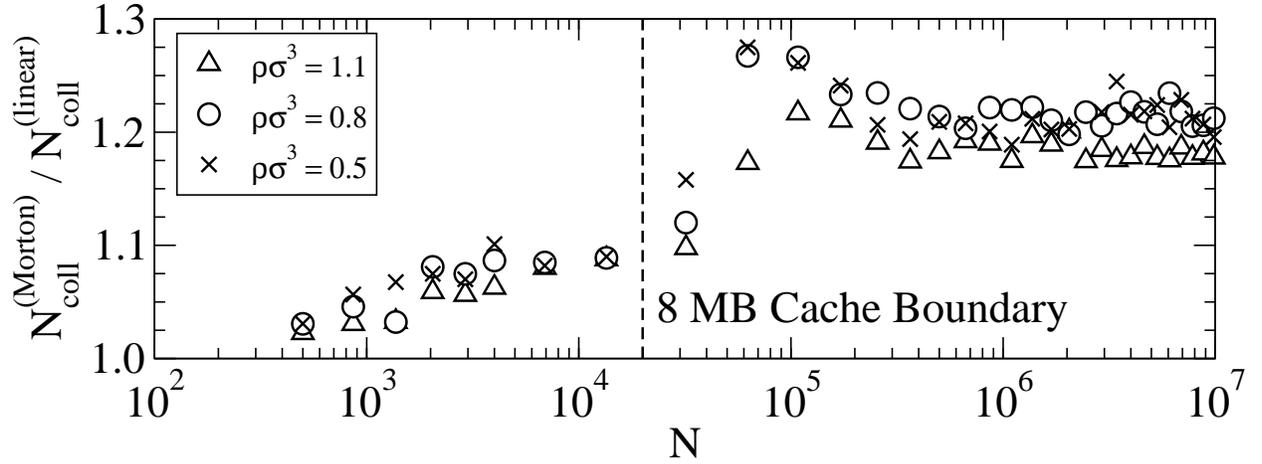}%
\caption{\label{fig:mortoncomparison}
The number of events processed per second using linear and Morton
ordered lists, $N_{\rm coll}^{\rm (linear)}$ and $N_{\rm coll}^{\rm
  (Morton)}$, respectively.  The dashed lines and system are described
in Fig.~\ref{fig:scaling}.}
\end{center}
\end{figure}

A comparison between Morton ordering and the typical linear ordering
is presented in Fig.~\ref{fig:mortoncomparison}.  Even inside the 8~MB
cache limit, the Morton ordering has a positive effect. This may be
due to localisation of memory accesses in the smaller L1 cache. For
more practical system sizes operating outside the cache, Morton
ordering offers a 10--28\% increase in event processing speed. This is
remarkable as only a single unsigned integer per cell is actually
optimized using this technique, proving its utility even in
event-driven simulations. At higher densities, the Morton ordering has
a slightly reduced effect as the ratio of particle to neighbor list
memory accesses is increased.  Overall, Morton ordering appears to be
an effective method of increasing the computational speed by reducing
the number of cache misses within a simulation.

The large effect of caching on the performance of the simulation
indicates at least half the time of simulations outside the cache
boundary are spent waiting on memory accesses. Cache simulations have
been performed using Callgrind~\cite{Weidendorfer_etal_2004}, and for
a density of $\rho\sigma^3=0.5$, approximately half of the cache
misses result from accesses to the contents of the neighbor lists.
The remaining cache misses are associated with accesses to particle
and event data.

\begin{figure*}[tb]
\begin{center}
  \includegraphics[width=\columnwidth,clip]{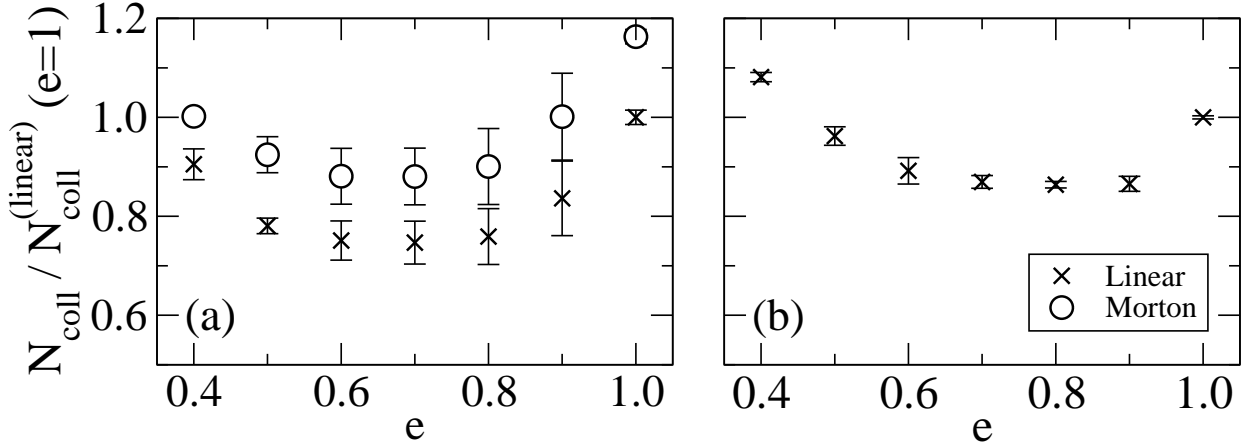}%
\caption{\label{fig:inelasScaling}
  The speedup factor versus density for (a) cooling and (b) sheared
  inelastic systems with $N=10^5$. The speedup factor is relative to the
  linear-neighbor-list elastic system. }
\end{center}
\end{figure*}

The code is also benchmarked for some more complex systems. The
collision model is extended to inelastic hard spheres with elasticity
$e$. In cooling simulations, the system is bounded by standard
periodic boundary conditions, and the temperature is rescaled to unity
every $2\times 10^6$ collisions. For sheared systems, Lees-Edwards
boundary conditions are used and rescaling is not required. Inelastic
particles ($e<1$) tend to cluster (see Fig.~\ref{fig:CoilSnapshot})
which increases the cost of updating the event list after a collision
(see Fig~\ref{fig:inelasScaling}a). However, at very low
inelasticities, the event processing rate increases as events become
correlated to ``rattling'' particles, improving the cache's
effectiveness and reducing the effect of Morton-ordered
neighbor-lists. This behavior is also apparent in the sheared
inelastic simulations (see Fig~\ref{fig:inelasScaling}b). Morton
ordering is quite effective in inelastic simulations, with a slight
enhancement over elastic simulations due to the increased spatial
correlations in the system. Morton ordering could not be tested in the
sheared system as it has not yet been ported to the sheared
neighbor-list.

\begin{figure}
\centering
\includegraphics[clip,width=0.9\columnwidth]{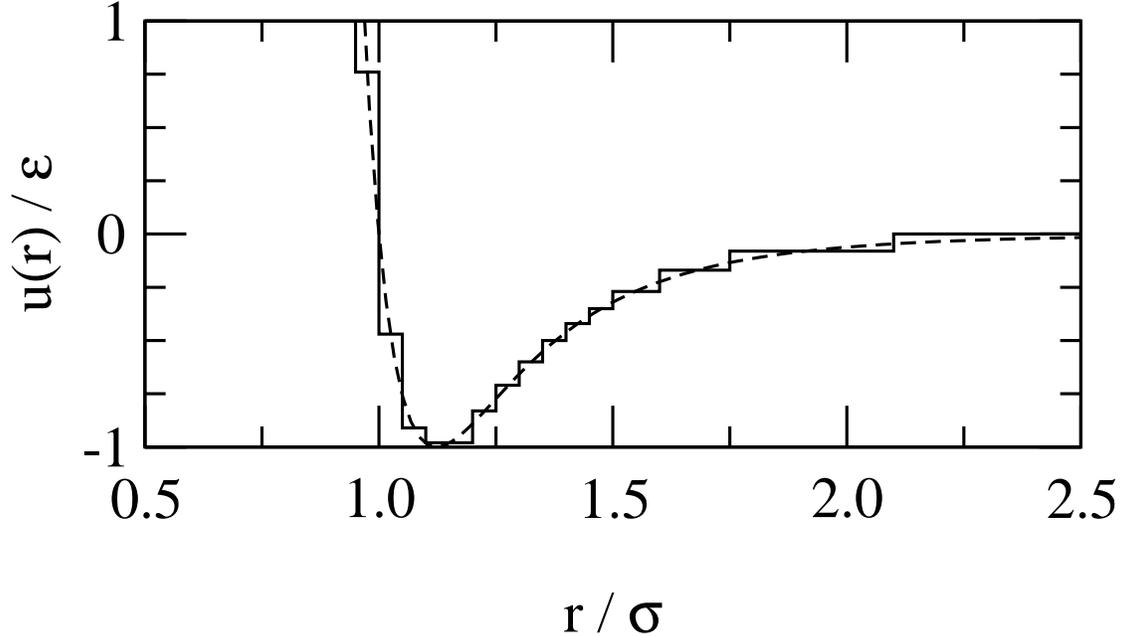}
\caption{\label{fig:LJstepped}
  A stepped potential (solid line) that approximates the Lennard-Jones
  potential (dashed line) (potential 6 of
  Ref.~\cite{Chapela_etal_1989}).}
\end{figure}

Finally, in order to assess the relative performance of an
event-driven algorithm with a time-stepping algorithm, simulations are
performed for the stepped and continuous variants of the Lennard-Jones
potential~\cite{Chapela_etal_1989} (see Fig.~\ref{fig:LJstepped}).
This potential provides a reasonable approximation to the
Lennard-Jones fluid and demonstrates the applicability of discrete
potentials to simple fluids. The continuous potential is simulated
using GROMACS 4.5.4~\cite{gromacs_4.0}, a popular and highly optimized
time stepping molecular dynamics package.  This comparison is biased
in favor of the time-stepping algorithm due to the shape of the
potential and the relative maturity of the GROMACS code; however, it
should be noted that, unlike time-stepping MD, event-driven algorithms
do not use a numerical integration scheme and are accurate to the
numerical precision of the machine. Both simulations consist of
$N=13500$ Lennard-Jones atoms with mass $m$, run for a simulation
length of $t=50(m\sigma^2/\epsilon)^{1/2}$, and using double precision
calculations.  The GROMACS simulations used the velocity Verlet
integrator, Verlet lists, a reduced time step of
$\Delta{}t=0.005(m\sigma^2/\epsilon)^{1/2}$, and a reduced cutoff
distance of $3\sigma$.

\begin{figure}
\centering
\includegraphics[clip,width=0.9\columnwidth]{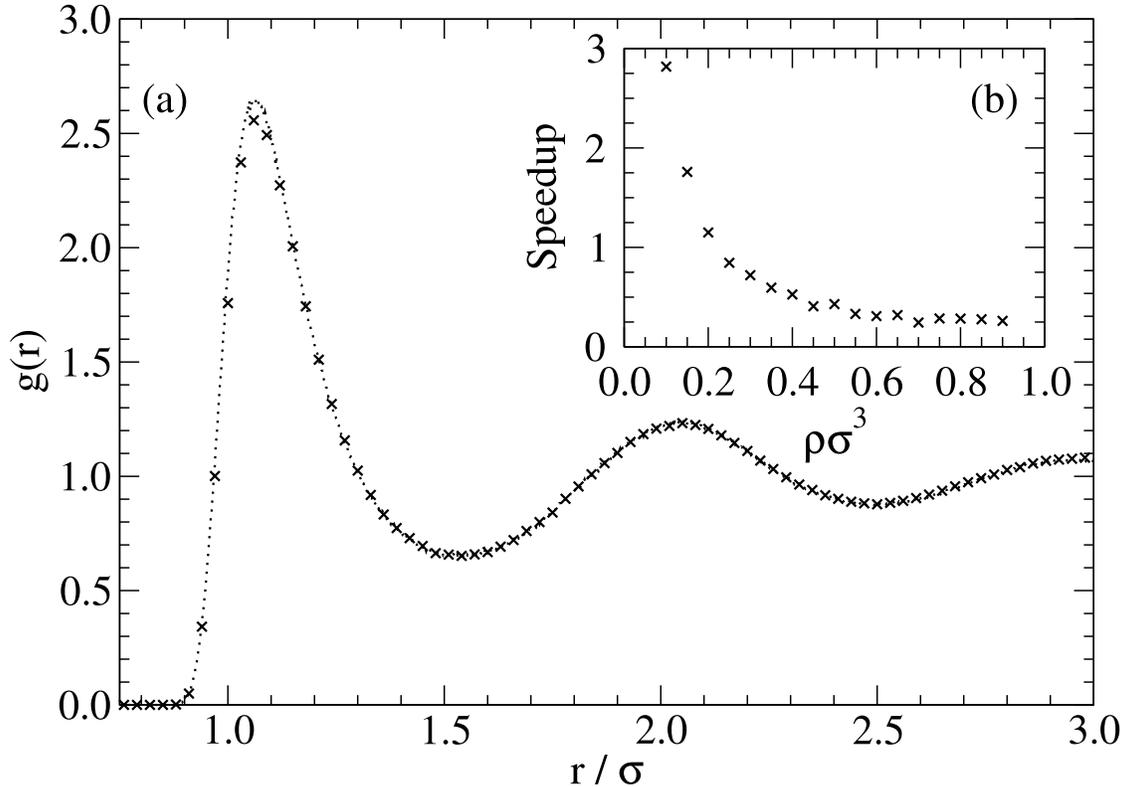}
\caption{\label{fig:gromacs}
  A comparison of (a) the radial distribution function $g(r)$ of a
  Lennard-Jones fluid at a number density of $\rho\sigma^3=0.85$ and a
  temperature of $k_BT/\varepsilon=1.3$ calculated using GROMACS
  (dotted line) and using DynamO (crosses) and (b) the speed of the
  DynamO simulations relative to the GROMACS simulations at
  $k_BT/\varepsilon=1.3$ over a range of densities. }
\end{figure}

A comparison of the calculated radial distribution functions and the
relative speed of the simulators are presented in
Fig.~\ref{fig:gromacs}. The radial distribution functions are in close
agreement, with a slight underestimation in the DynamO results at a
distance of $r/\sigma\approx1.1$. At low densities, DynamO
significantly outperforms GROMACS as expected as event-driven dynamics
is optimal in collisional regimes.  Surprisingly, DynamO also performs
well into the liquid phase, with GROMACS only displaying a $4\times$
speedup. This is a small cost when it is considered that the
event-driven algorithm solves its dynamics without truncation error.

\section{Conclusions
\label{sec:conclusions}}

We have detailed the fundamental components of DynamO, a modern
event-driven molecular-dynamics simulation package.  The program is
distributed under the GNU General Public License.
The full source code and documentation are freely available online at
{\tt http://www.marcusbannerman.co.uk/dynamo}.
The program provides reference implements for many modern algorithms
for event-driven simulations and also includes several new techniques
for mitigating round-off error, improving speed, and optimizing memory
access patterns.  The latter is achieved by preserve cache locality
through using Morton ordering to store neighbor entries in spatially
localized clusters.  The speed of accessing memory appears to be a
significant bottleneck in simulating systems with simple potentials.
%
Through benchmark simulations on single component elastic-hard-sphere
systems, we have demonstrated that DynamO exhibits an ${\mathcal
  O}(1)$ scaling with system size of the computational cost of
executing events.  This leads to an overall scaling of ${\mathcal
  O}(N)$ for a set duration of simulation time.  This allows the rapid
simulation of both complex and large ($10^7$ particle/atom) systems
while extracting the long-time behavior.

Many systems can be explored with the package in its current state;
however, there are a few planned extensions which will bring the
package to the level of generality of modern soft potential packages.

Stepped potentials~\cite{Chapela_etal_1989} are already available in
DynamO, allowing the straightforward approximation of rotationally
symmetric soft potentials. Asymmetric potential dynamics are
significantly more complex, especially in the case of hard
particles~\cite{Donev_etal_2005}. The determination of the time to
collision requires considerable care, and these algorithms are often
specialized to the underlying
potential~\cite{Frenkel_Maguire_1983}. On the other hand, soft
potential dynamics are widely used for modelling due to the relative
ease with which new potentials can be implemented. A natural step
forward for event-driven dynamics is in the implementation of the
framework developed by van~Zon and
Schofield~\cite{vanZon_Schofield_2008}, which generalizes the
implementation of asymmetric discrete models by using a soft potential
to generate and solve the dynamics of an equivalent ``terraced''
potential. A partial implementation is already available, although care
must still be taken in the discretization of the soft-potential;
further research is needed to develop this technique.

Long-ranged potentials, such as those due to electrostatic
interactions, do not yet have an event-driven equivalent.  The
implementation of a stepped force field is not difficult; however,
coupling the particle positions to the field is not
trivial~\cite{Sigurgeirsson_2001}. The detection of events becomes
prohibitively expensive due to the added complexity of the free flight
phase and typically only exists to ensure that an underlying time
stepping integration does not fail by incorrectly generating
overlapping hard particles~\cite{Sigurgeirsson_2001}. In the limit of
a weak and long-ranged field within a large system, this coupling
might be implemented as a boundary condition of a stepped potential
grid.  These techniques must be developed before event-driven dynamics
can be effectively utilized in modeling charged systems.



Event-driven simulations are serial by nature; however, attempts have
been made to develop parallel
algorithms~\cite{Marin_1997,Miller_Luding_2004}.  The simulations are
split into computational cells and divided among a collection of
processors. Each cell is then run independently of all others until an
event occurs at the boundaries of the cells, forcing a
synchronization. As the number of processors increases, the
synchronization events become more frequent and will limit the
scalability; however, excellent performance has been demonstrated for up
to 128 processors~\cite{Miller_Luding_2004}. At the core of these
parallel algorithms is a serial implementation, and DynamO has already
been used to simulate systems of 32 million particles on a single
processor. Yet, parallel computation will be required as the complexity
of the underlying potentials increases.  Further developments are
needed to optimize memory usage and to explore the possibility of
parallelizing the algorithm within a computational cell.

\appendix
\section{Program design and Features
\label{sec:programdesign}}

The development of DynamO has focused on generating a flexible,
modular simulator where systems can be constructed from an array of
available interactions, conditions, and dynamics.  DynamO is written
in \Cpp using an object orientated design.  This helps ensure that the
code is both extensible and maintainable, provided the classes have
well defined interfaces and tasks.  All input and output files are in
XML to allow easy generation and alteration of system conditions.
The implementation of DynamO utilizes only free, open source
libraries, including the BOOST ({\tt www.boost.org}) libraries.

DynamO was originally written to perform NVE molecular-dynamics
simulations of particles interacting through spherically symmetric,
discrete potentials.  However, because of its flexible design, DynamO
has been extended to perform a wide variety of calculations for
several different types of systems.
The dynamics of infinitely thin lines~\cite{Frenkel_Maguire_1983} has
already been incorporated within DynamO, and other shapes can also be
included.  Constant temperature simulations are performed through the
use of the Andersen thermostat~\cite{Andersen_1980}.  Multiple
simulations can be executed concurrently and combined with the replica
exchange method~\cite{Swendsen_Wang_1986} to expedite the
equilibration of systems with rough energy landscapes.  Umbrella
potentials can also be applied to sample specific regions of the phase
space of a system.  Finally, stepped potentials can be used to
approximate rotationally symmetric soft potentials.

Alternate dynamics can also be easily implemented within DynamO.  For
example, compression dynamics, where the particles in the system grow
with time, is already implemented.  DynamO is also used
\cite{Bannerman_etal_2009,Bannerman_Lue_2009} as a framework to
perform direct simulation Monte Carlo calculations for the Enskog
equation.

Simple code hierarchies have been suggested
previously~\cite{Erleben_2002}; however, the level of complexity of
modern simulations requires a finer grained class structure than
previously outlined.

\begin{figure*}[tb]
\centering
\includegraphics[clip,width=0.75\textwidth]{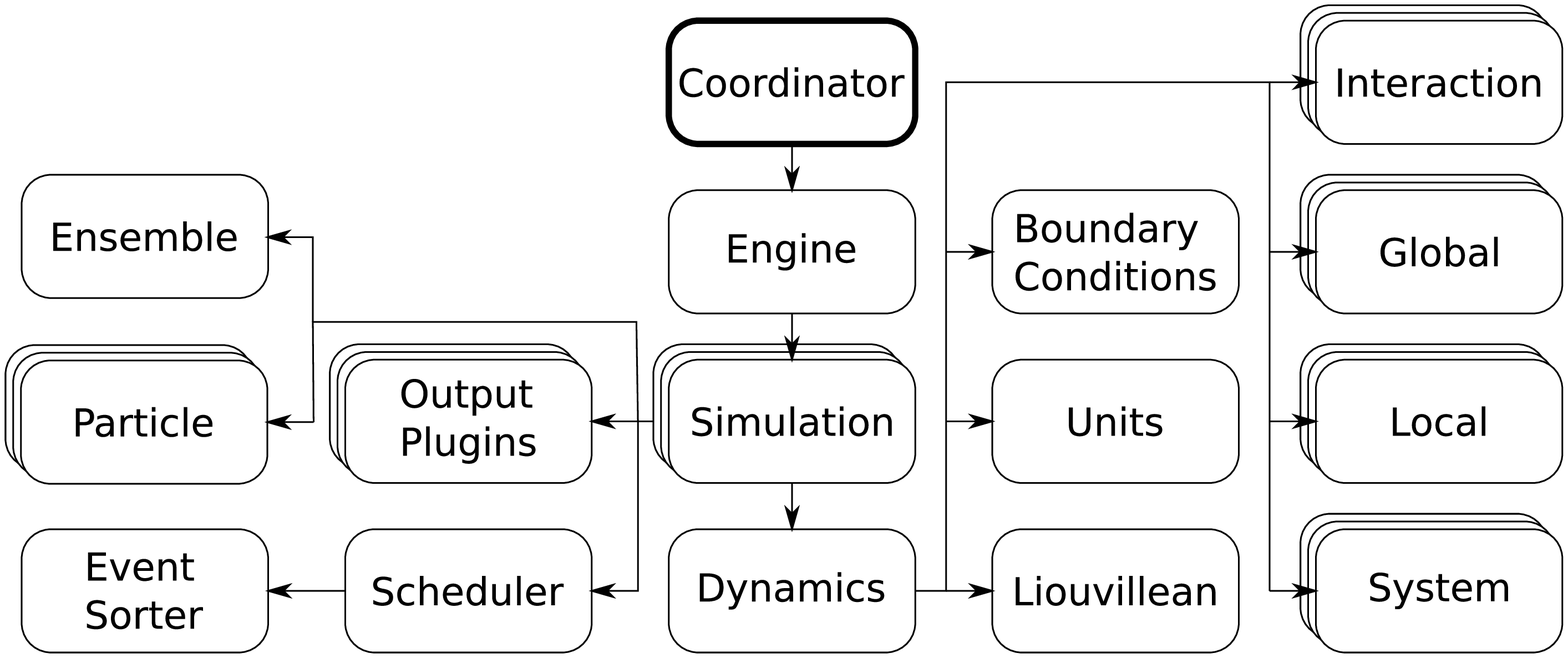}
\caption{\label{fig:progstruct}
The class hierarchy of DynamO.  Only the classes key to the algorithm
are displayed.  Arrows indicate the nesting of classes, and stacked
boxes indicate multiple instances of the class may occur.}
\end{figure*}

The class hierarchy of DynamO is presented in
Fig.~\ref{fig:progstruct}.  Typically each class has several
implementations which are selected at run time through the input
files.
Below, we detail the scope of each class, along with the currently
implemented features:

\begin{list}{}{\settowidth{\leftmargin}{}\setlength{\itemsep}{0pt}}
\item {\bf Coordinator:} {\em Abstracts the user interface and system
    calls.} \\%
  This class encapsulates the entire program and provides the user
  interface. The initialization of operating system features, such as
  threading, is also performed here.
\item {\bf Engine:} {\em Organizes a collection of simulations to
    achieve a task.}\\%
  In its simplest form, a single simulation is run to obtain output;
  however, replica exchange techniques~\cite{Swendsen_Wang_1986} and a
  method to perform isotropic compression of
  configurations~\cite{Woodcock_1981} are also available. The replica
  exchange technique runs several simulations in parallel with a
  Monte Carlo move to increase the ergodicity of low temperature
  trajectories.
\item {\bf Simulation:} {\em Encapsulates a single simulation.}\\%
  This class represents a single simulation, containing an array of
  particles, classes describing the dynamics, and data collection
  classes.  The primary function of the simulation class is to
  initialize and maintain these classes in evolving the system through
  time.
\item {\bf Scheduler:} {\em Maintains the list of future events.}\\%
  This class is responsible for executing events and maintaining the
  FEL. Several variants exist, a ``dumb'' scheduler, a scheduler
  capable of interfacing with a specialized Global that implements a
  neighbor list, and a multi-threaded neighbor-list scheduler.
\item {\bf Event Sorter:} {\em Sorts events in the FEL.} \\%
  Provides a method of sorting the FEL, which contains an array of
  PELs. A complete binary tree~\cite{Marin_Cordero_1995} and a
  bounded priority queue~\cite{Paul_2007} are implemented. 
\item {\bf Particle:} {\em Container for single particle data.} \\%
  This class encapsulates the minimal single particle data. This
  includes the particle ID number, position, velocity, and local time.
\item {\bf Output Plugins:} {\em Data collection routines.}  \\%
  A wide range of plugins are available, including radial distribution
  functions, the complete set of Green-Kubo expressions for mixtures,
  and visualization plugins for VMD ({\tt
    www.ks.uiuc.edu/Research/vmd}), Povray ({\tt www.povray.org}), and
  Geomview ({\tt www.geomview.org}).
\item {\bf Ensemble:} {\em Describes the ensemble of the
    simulation.}\\%
  This is used to ensure the simulation ensemble is valid for certain
  output plugins and for replica exchange.
\item {\bf Dynamics:} {\em Encapsulates the dynamics methods of the
    system.}\\%
  The dynamics class initializes and maintains classes relating to the
  dynamics of the system. The actual dynamics are implemented in
  classes contained within this class.
\item {\bf Units:} {\em Provides functions to scale between simulation
    and input units.} \\%
  Simulations are typically optimal in a unit set other than the input
  settings.  For example, if the dimensions of the simulation box are
  scaled to one then the enforcement of the periodic boundary
  conditions reduces to a rounding operation~\cite{Haile_1997}.
\item {\bf Liouvillean:} {\em Contains simple functions to describe
    the evolution of the system.}\\%
  This class implements event testing for basic shapes (e.g., spheres,
  lines, planes), particle evolution and event dynamics. These are
  then used by interactions, locals and globals to implement an event.
  Several implementations exist, including Newtonian, isotropic
  compression, Enskog direct simulation Monte Carlo (DSMC) in Sllod
  coordinates~\cite{Hopkins_Shen_1992}, and axisymmetric rotational
  dynamics.
\item {\bf Boundary Conditions:} {\em Specifies the limits of the
    simulation box.}\\%
  Square, rectangular, periodic, sliding brick
  shearing~\cite{Lees_Edwards_1972}, and infinite boundary conditions
  are currently implemented.
\item {\bf Interaction:} {\em Two particle events.}\\%
  Interactions between two particles are derived from this class.
  Many interactions have already been implemented (e.g., stepped
  potentials, hard spheres, parallel cubes, square wells, square-well
  bonds~\cite{SMITH_ETAL_1997}, thin
  needles~\cite{Frenkel_Maguire_1983}, and an optimized square-well
  sequenced-polymer interaction).
\item {\bf Global:} {\em Single particle events.} \\%
  Events which affect only one particle, regardless of its position in
  the simulation are derived from this class. This is primarily used
  for neighbor lists, specializations include overlapping
  cells~\cite{Krantz_1996}, mixed
  methods~\cite{Vrabecz_Toacuteth_2006}, cells in shearing, Morton
  ordered neighbor lists, and a sentinel event to ensure the nearest
  image condition does not result in invalid dynamics at low
  density/small system size~\cite{Frenkel_Maguire_1983}.
\item {\bf Local:} {\em Single particle event, localized in space.}\\%
  These events only effect a region of space and are inserted into the
  neighbor lists when available. This is to reduce the number of event
  tests required.  Solid, thermostatted~\cite{Poschel_Schwager_2005}
  and oscillating planar/cylindrical/spherical walls are implemented
  here.
\item {\bf System:} {\em Simulation and multiple-particle events.}\\%
  Any type of event which is not compatible with the definition of a
  Local, Global or Interaction is implemented here.  Includes DSMC
  interactions~\cite{bird_1994}, Andersen
  thermostats~\cite{Andersen_1980}, and simulation termination
  conditions.  Umbrella potentials are also implemented here as they
  are many-body interactions.
\end{list}

\acknowledgments{MN Bannerman and R Sargant acknowledge support from the EPSRC DTA.}


\begin{thebibliography}{10}

\bibitem{Fish_2006}
Fish, J., {\em J. Nanoparticle Res.}, 2006, {\bf 8}, 577--594.

\bibitem{Hansson_etal_2002}
Hansson, T.; Oostenbrink, C. and van Gunsteren, W., {\em Current Opinion in
  Structural Biology}, 2002, {\bf 12}(2), 190 -- 196.

\bibitem{Wong_McCammon_2003}
Wong, C.~F. and McCammon, J.~A. In {\em Protein Simulations}, Daggett, V., Ed.,
  Vol. ~66 of {\em Advances in Protein Chemistry;}
\newblock Academic Press, 2003;
\newblock pages 87 -- 121.

\bibitem{Jorgensen_2004}
Jorgensen, W.~L., {\em Science}, 2004, {\bf 303}, 1813--1818.

\bibitem{Rahman_1964}
Rahman, A., {\em Phys. Rev.}, 1964, {\bf 136}, A405--A411.

\bibitem{Verlet_1967}
Verlet, L., {\em Phys. Rev.}, 1967, {\bf 159}, 98.

\bibitem{Plimpton_1995}
Plimpton, S., {\em J. Comput. Phys.}, 1995, {\bf 117}, 1--19.

\bibitem{gromacs_2005}
Spoel, D.~V.~D.; Lindahl, E.; Hess, B.; Groenhof, G.; Mark, A.~E. and
  Berendsen, H.~J.~C., {\em J. Comput. Chem.}, 2005, {\bf 26}, 1701--1718.

\bibitem{gromacs_4.0}
Hess, B.; Kutzner, C.; Van Der~Spoel, D. and Lindahl, E., {\em J. Chem. Theory
  Comput.}, 2008, {\bf 4}(3), 435--447.

\bibitem{Phillips_etal_2005}
Phillips, J.~C.; Braun, R.; Wang, W.; Gumbart, J.; Tajkhorshid, E.; Villa, E.;
  Chipot, C.; Skeel, R.~D.; Kal\'{e}, L. and Schulten, K., {\em J. Comput.
  Chem.}, 2005, {\bf 26}(16), 1781--1802.

\bibitem{Bowers_etal_2006}
Bowers, K.~J.; Chow, E.; Xu, H.; Dror, R.~O.; Eastwood, M.~P.; Gregersen,
  B.~A.; Klepeis, J.~L.; Kolossv\'{a}ry, I.; Moraes, M.~A.; Sacerdoti, F.~D.;
  Salmon, J.~K.; Shan, Y. and Shaw, D.~E. In {\em Proceedings of the ACM/IEEE
  Conference on Supercomputing (SC06)}, Tampa, Florida, 2006.

\bibitem{Limbach_etal_2006}
Limbach, H.-J.; Arnold, A.; Mann, B.~A. and Holm, C., {\em Comput. Phys.
  Commun.}, 2006, {\bf 174}(9), 704--727.

\bibitem{Cheatham_Young_2001}
Cheatham~III, T.~E. and Young, M., {\em Biopolymers}, 2001, {\bf 56}, 232--256.

\bibitem{Ponder_Case_2003}
Ponder, J. and Case, D., {\em Adv. Prot. Chem.}, 2003, {\bf 66}, 27--85.

\bibitem{CHARMMFF}
MacKerel~Jr., A.; Brooks~III, C.; Nilsson, L.; Roux, B.; Won, Y. and Karplus,
  M., In {\em The Encyclopedia of Computational Chemistry}, von R.~Schleyer,
  P., Ed., John Wiley \& Sons: Chichester, 1998;
\newblock Vol. ~1;
\newblock pages 271--277.

\bibitem{Cundall_Strack_1979}
Cundall, P.~A. and Strack, O.~D.~L., {\em G\'eotechnique}, 1979, {\bf 29},
  47--65.

\bibitem{Brillantov_etal_1996}
Brilliantov, N.; Spahn, F.; Hertzsch, J.-M. and P\"oschel, T., {\em Phys. Rev.
  E}, 1996, {\bf 53}, 5382--5392.

\bibitem{Deltour_Barrat_1997}
Deltour, P. and Barrat, J.~L., {\em Journal De Physique I}, 1997, {\bf 7}(1),
  137--151.

\bibitem{Haff_Werner_1986}
Haff, P.~K. and Werner, B.~T., {\em Powder Tech.}, 1986, {\bf 48}, 239--245.

\bibitem{Walton_Braun_1986a}
Walton, O.~R. and Braun, R.~L., {\em J. Rheol.}, 1986, {\bf 30}, 949.

\bibitem{Chapela_etal_1989}
Chapela, G.; Scriven, L.~E. and Davis, H.~T., {\em J. Chem. Phys.}, 1989, {\bf
  91}, 4307.

\bibitem{Elliot_2002}
Elliott, J.~R., {\em Fluid Phase Equilibria}, 2002, {\bf 194--197}, 161--168.

\bibitem{Molinero_Goddard_2004}
Molinero, V. and Goddard, W.~A., {\em J. Phys. Chem. B}, 2004, {\bf 108}(4),
  1414--1427.

\bibitem{Marrink_etal_2007}
Marrink, S.~J.; Risselada, H.~J.; Yefimov, S.; Tieleman, D.~P. and de~Vries,
  A.~H., {\em J. Phys. Chem. B}, 2007, {\bf 111}(27), 7812--7824.

\bibitem{Monticelli_etal_2008}
Monticelli, L.; Kandasamy, S.~K.; Periole, X.; Larson, R.~G.; Tieleman, D.~P.
  and Marrink, S.-J., {\em J. Chem. Theory Comput.}, 2008, {\bf 4}(5),
  819--834.

\bibitem{ALDER_WAINWRIGHT_1957}
Alder, B.~J. and Wainwright, T.~E., {\em J. Chem. Phys.}, 1957, {\bf 27}(5),
  1208--1209.

\bibitem{Goldhirsch_Zanetti_1993}
Goldhirsch, I. and Zanetti, G., {\em Phys. Rev. Lett.}, 1993, {\bf 70},
  1619--1622.

\bibitem{Unlu_etal_2004}
Unlu, O.; Gray, N.~H.; Gerek, Z.~N. and Elliott, J.~R., {\em Ind. Eng. Chem.
  Res.}, 2004, {\bf 43}(7), 1788--1793.

\bibitem{Elliott_etal_2007}
Elliott, J.~R.; Vahid, A. and Sans, A.~D., {\em Fluid Phase Equilibr.}, 2007,
  {\bf 256}(1-2), 4 -- 13.

\bibitem{Vahid_Elliott_2010}
Vahid, A. and Elliott, J.~R., {\em AIChE J.}, 2010, {\bf 56}(2), 485--505.

\bibitem{Woodhead_Hall_2010}
Woodhead, J.~L. and Hall, C.~K., {\em Langmuir}, 2010, {\bf 26}(19),
  15135--15141.

\bibitem{Schultz_etal_2002}
Schultz, A.~J.; Hall, C.~K. and Genzer, J., {\em J. Chem. Phys.}, 2002, {\bf
  117}(22), 10329--10338.

\bibitem{Nguyen_Hall_2004}
Nguyen, H.~D. and Hall, C.~K., {\em Biophys. J.}, 2004, {\bf 87}, 4122--4134.

\bibitem{Hall_Wagoner_2006}
Hall, C.~K. and Wagoner, V.~A. In {\em Amyloid, Prions, and Other Protein
  Aggregates, Part B}, Kheterpal, I. and Wetzel, R., Eds., Vol.  412 of {\em
  Methods in Enzymology;}
\newblock Academic Press, 2006;
\newblock pages 338 -- 365.

\bibitem{Marchut_Hall_2007}
Marchut, A.~J. and Hall, C.~K., {\em Proteins: Struct., Funct., Bioinf.}, 2007,
  {\bf 66}(1), 96--109.

\bibitem{Cheon_etal_2010}
Cheon, M.; Chang, I. and Hall, C.~K., {\em Proteins: Struct., Funct., Bioinf.},
  2010, {\bf 78}(14), 2950--2960.

\bibitem{Allen_etal_1989}
Allen, M.~P.; Frenkel, D. and Talbot, J., {\em Comp. Phys. Rep.}, 1989, {\bf
  9}(6), 302--353.

\bibitem{ALDER_WAINWRIGHT_1959}
Alder, B.~J. and Wainwright, T.~E., {\em J. Chem. Phys.}, 1959, {\bf 31}(2),
  459--466.

\bibitem{Rapaport_1980}
Rapaport, D.~C., {\em J. Comput. Phys.}, 1980, {\bf 34}(2), 184--201.

\bibitem{Lubachevsky_1991}
Lubachevsky, B.~D., {\em Int. J. Comput. Phys.}, 1991, {\bf 94}, 255--283.

\bibitem{Marin_etal_1993}
Marin, M.; Risso, D. and Cordero, P., {\em J. Comput. Phys.}, 1993, {\bf 109},
  306--317.

\bibitem{Sigurgeirsson_2001}
Sigurgeirsson, H.; Stuart, A. and Wan, W.-L., {\em J. Comput. Phys.}, 2001,
  {\bf 172}, 766--807.

\bibitem{Donev_2009}
Donev, A., {\em Simulation}, 2009, {\bf 85}(4), 229--242.

\bibitem{Bannerman_etal_2010}
Bannerman, M.~N.; Lue, L. and Woodcock, L.~V., {\em J. Chem. Phys.}, 2010, {\bf
  132}, 084507.

\bibitem{Bannerman_etal_2009}
Bannerman, M.~N.; Green, T.~E.; Grassia, P. and Lue, L., {\em Phys. Rev. E},
  2009, {\bf 79}, 041308.

\bibitem{Bannerman_etal_2011}
Bannerman, M.~N.; Kollmer, J.~E.; Sack, A.; Heckel, M.; M\"uller, P. and
  P\"oschel, T., {\em Phys. Rev. E}, 2011, {\bf 84}, 011301.

\bibitem{Bannerman_Lue_2010}
Bannerman, M.~N. and Lue, L., {\em J. Chem. Phys.}, 2010, {\bf 133}, 124506.

\bibitem{Bannerman_Lue_2009}
Bannerman, M.~N. and Lue, L., {\em J. Chem. Phys.}, 2009, {\bf 130}, 164507.

\bibitem{Hoover_etal_2009}
Hoover, W.~G.; Hoover, C.~G. and Bannerman, M.~N., {\em J. Stat. Phys.}, 2009,
  {\bf 136}, 715--732.

\bibitem{Bannerman_etal_2009_2}
Bannerman, M.~N.; Magee, J. and Lue, L., {\em Phys. Rev. E}, 2009, {\bf 80},
  021801.

\bibitem{Marin_Cordero_1996}
Marin, M. and Cordero, P. In Borcherds, P. and Bubak, M., Eds., {\em 8th Joint
  EPS-APS International Conference on Physics Computing}, pages 315--318. World
  Scientific, 1996.

\bibitem{Krantz_1996}
Krantz, A.~T., {\em TOMACS}, 1996, {\bf 6}(3), 185--209.

\bibitem{Vrabecz_Toacuteth_2006}
Vrabecz, A. and T\'oth, G., {\em Mol. Phys.}, 2006, {\bf 104}(12), 1843--1853.

\bibitem{Jefferson_1985}
Jefferson, D.~R., {\em TOPLAS}, 1985, {\bf 7}(3), 404--425.

\bibitem{Marin_Cordero_1995}
Marin, M. and Cordero, P., {\em Comp. Phys. Comm.}, 1995, {\bf 92}(2--3),
  214--224.

\bibitem{Paul_2007}
Paul, G., {\em J. Comput. Phys.}, 2007, {\bf 221}(2), 615--625.

\bibitem{ALLEN_TILDESLEY_1987}
Allen, M.~P. and Tildesley, D.~J., {\em Computer simulation of liquids}, Oxford
  University Press, Bristol, 1987.

\bibitem{Raman_Wise_2008}
Raman, R. and Wise, D.~S., {\em IEEE Trans. Comp.}, 2008, {\bf 57}(4),
  567--573.

\bibitem{Adams_Wise_2006}
Adams, M.~D. and Wise, D.~S., {\em ACM SIGPLAN Notices}, 2006, {\bf 41}(5),
  39--45.

\bibitem{Poschel_Schwager_2005}
P\"oschel, T. and Schwager, T., {\em Computational Granular Dynamics},
  Springer, New York, 2005.

\bibitem{Atkinson_etal_1986}
Atkinson, M.~D.; Sack, J.-R.; Santoro, N. and Strothotte, T., {\em
  Communications of the ACM}, 1986, {\bf 29}(10), 996--1000.

\bibitem{Andersen_1980}
Andersen, H.~C., {\em J. Chem. Phys.}, 1980, {\bf 72}, 2384.

\bibitem{Weidendorfer_etal_2004}
Weidendorfer, J.; Kowarschik, M. and Trinitis, C. In {\em Computational Science
  - ICCS 2004}, Vol.  3038, pages 440--447, Berlin, 2004. Springer-Verlag.

\bibitem{Donev_etal_2005}
Donev, A.; Torquato, S. and Stillinger, F., {\em J. Comput. Phys.}, 2005, {\bf
  202}, 737--764.

\bibitem{Frenkel_Maguire_1983}
Frenkel, D. and Maguire, J.~F., {\em Mol. Phys.}, 1983, {\bf 49}(3), 503--541.

\bibitem{vanZon_Schofield_2008}
van Zon, R. and Schofield, J., {\em J. Chem. Phys.}, 2008, {\bf 128}, 154119.

\bibitem{Marin_1997}
Marin, M., {\em Comp. Phys. Comm.}, 1997, {\bf 102}, 81--96.

\bibitem{Miller_Luding_2004}
Miller, S. and Luding, S., {\em J. Comput. Phys.}, 2004, {\bf 193}(1),
  306--316.

\bibitem{Swendsen_Wang_1986}
Swendsen, R.~H. and Wang, J., {\em Phys. Rev. Lett.}, 1986, {\bf 57},
  2607--2609.

\bibitem{Erleben_2002}
Erleben, K. Module based design for rigid body simulators Technical report,
  University of Copenhagen, 2002.

\bibitem{Woodcock_1981}
Woodcock, L.~V., {\em Ann. NY Acad. Sci.}, 1981, {\bf 371}(OCT), 274--298.

\bibitem{Haile_1997}
Haile, J.~M., {\em Molecular Dynamics Simulation - Elementary Methods},
  Wiley-Interscience, New York, 1997.

\bibitem{Hopkins_Shen_1992}
Hopkins, M.~A. and Shen, H.~H., {\em J. Fluid Mech.}, 1992, {\bf 244},
  477--491.

\bibitem{Lees_Edwards_1972}
Lees, A.~W. and Edwards, S.~F., {\em J. Phys. C}, 1972, {\bf 5}, 1921--1929.

\bibitem{SMITH_ETAL_1997}
Smith, S.~W.; Hall, C.~K. and Freeman, B.~D., {\em J. Comput. Phys.}, 1997,
  {\bf 134}(1), 16--30.

\bibitem{bird_1994}
Bird, G.~A., {\em Molecular gas dynamics and the direct simulation of gas
  flows}, Oxford University Press, Oxford, 1994.

\end{thebibliography}

\clearpage
\section*{Figure captions}

\newcounter{fig}
\begin{list}{\bf FIG.~\arabic{fig}}
  {\usecounter{fig} \leftmargin 2cm \labelwidth 1.5cm \labelsep 0.5cm}

\item A two dimensional $4\times4$ array stored
  in row major order and Morton order in memory.

\item  The speedup and memory cost of using a MinMax PEL of various sizes
  relative to storing only a single event per particle. The dashed
  lines are the values for using a STL priority\_queue. The results
  are for a monodisperse system of $N=10^5$ elastic hard spheres
  simulated on an Intel\textregistered\ Core\texttrademark i5 750 with
  8~MB L3 cache and 8~GB of RAM.

\item  A snapshot of $5\times10^5$ sheared inelastic hard spheres,
  generated live from a simulation, displaying a characteristic
  clustering behavior of granular systems. On a NVIDIA\textregistered\ 
  Geforce GTX 260\texttrademark, the image is updated at 15 frames per
  second.

\item  The number of collision events processed per second using linear
  neighbor lists $N_{\rm coll}^{(\rm linear)}$ as a function of (a)
  the number of particles $N$ and (b) the number density $\rho$.  The
  lines indicating the cache memory boundary is approximate as
  different densities incur slightly different memory requirements.
  The results are for a monodisperse system of elastic hard spheres
  simulated on an Intel\textregistered Core\texttrademark i5 750 with
  8~MB L3 cache and 8~GB of RAM.

\item The number of events processed per second using linear and Morton
ordered lists, $N_{\rm coll}^{\rm (linear)}$ and $N_{\rm coll}^{\rm
  (Morton)}$, respectively.  The dashed lines and system are described
in Fig.~\ref{fig:scaling}.

\item  The speedup factor versus density for (a) cooling and (b) sheared
  inelastic systems with $N=10^5$. The speedup factor is relative to the
  linear-neighbor-list elastic system.

\item  A stepped potential (solid line) that approximates the Lennard-Jones
  potential (dashed line) (potential 6 of
  Ref.~\cite{Chapela_etal_1989}).

\item  A comparison of (a) the radial distribution function $g(r)$ of a
  Lennard-Jones fluid at a number density of $\rho\sigma^3=0.85$ and a
  temperature of $k_BT/\varepsilon=1.3$ calculated using GROMACS
  (dotted line) and using DynamO (crosses) and (b) the speed of the
  DynamO simulations relative to the GROMACS simulations at
  $k_BT/\varepsilon=1.3$ over a range of densities.

\item The class hierarchy of DynamO.  Only the classes key to the algorithm
are displayed.  Arrows indicate the nesting of classes, and stacked
boxes indicate multiple instances of the class may occur.

\end{list}

\end{document}